\documentclass[twocolumn]{IEEEtran}

\IEEEoverridecommandlockouts
\usepackage{cite}
\usepackage{amsmath,amssymb,amsfonts}
\usepackage{algorithmic}
\usepackage{graphicx}
\usepackage{textcomp}
\usepackage{xcolor}
\usepackage{bm}
\usepackage{stfloats}
\usepackage{mathrsfs}
\usepackage{algorithm}
\usepackage{tabularx}

\def\BibTeX{{\rm B\kern-.05em{\sc i\kern-.025em b}\kern-.08em
    T\kern-.1667em\lower.7ex\hbox{E}\kern-.125emX}}
\begin{document}

\title{Rapid Phase Ambiguity Elimination Methods for DOA Estimator via Hybrid Massive MIMO Receive Array}

\author{Xichao Zhan, Yiwen Chen, Feng Shu, Xin Cheng, Yuanyuan Wu, Qi Zhang, Yifang Li, and Peng Zhang

\thanks{This work was supported in part by the National Natural Science Foundation of China (Nos. 62071234, 62071289, and 61972093), the Hainan Major Projects (ZDKJ2021022), the Scientific Research Fund Project of Hainan University under Grant KYQD(ZR)-21008 and KYQD(ZR)-21007, and the National Key R\&DProgram of China under Grant 2018YFB180110. (Corresponding authors: Feng Shu, Email: shufeng0101@163.com).}
\thanks{Xichao Zhan, Yiwen Chen, Qi Zhang, Peng Zhang, Yuanyuan Wu and Feng Shu are with the School of Information and Communication Engineering, Hainan University, Haikou, 570228, China.}
\thanks{Xin Cheng, Yifang Li and Feng Shu are with the School of Electronic and Optical Engineering, Nanjing University of Science and Technology, 210094, China.}


}

\maketitle

\begin{abstract}

For a sub-connected hybrid multiple-input multiple-output (MIMO) receiver with $K$ subarrays and $N$ antennas,  there exists a challenging problem  of how to rapidly remove phase ambiguity in only single time-slot. First, a DOA estimator of maximizing received power (Max-RP)  is proposed to find the maximum value of $K$-subarray output powers, where each subarray is in charge of one sector, and the center angle of the sector corresponding to the maximum output is the estimated true DOA. To make an enhancement on precision, Max-RP plus quadratic interpolation (Max-RP-QI) method is designed. In the proposed Max-RP-QI, a quadratic interpolation scheme is adopted to interpolate the three DOA values corresponding to the largest three receive powers of Max-RP. Finally, to achieve the CRLB, a Root-MUSIC plus Max-RP-QI scheme is developed. Simulation results show that the proposed three methods eliminate the phase ambiguity during one time-slot and also show low-computational-complexities. In particular, the proposed  Root-MUSIC plus Max-RP-QI scheme can reach the CRLB, and the proposed Max-RP and Max-RP-QI are still some performance losses $2dB\thicksim4dB$ compared to the CRLB.

\end{abstract}
\section{Introduction}
Due to the development of computing processors and digital signal processing technology, array signal processing (ASP), particularly large-scale ASP, has attracted more and more attention. As one of key techniques of ASP \cite{T.Engin-09, Handbook-19}, direction of arrival DOA  estimation  has been widely used in mobile communications \cite{ZhaoJunhui-5G}, direction modulation (DM) network \cite{SF-JSAC, SF-Access}, radar, sonar, earthquake monitoring, aerospace, mmWave communications \cite{MWA-2016}, unmanned aerial vehicle (UAV) communications \cite{ChengXin-TVT, ZhaoJunhui-UAV}, satellite communications, and angle of arrival (AOA) positioning \cite{Li-Yiwen-20-5g, WangYue-AOA-TDOA}.

It is critical to infer the presence or absence of an emitter before performing DOA measurements. If there exists no emitter, then it is obvious that no  DOA estimation operation is required. In \cite{ZhangRui-GLRT-approach}, two signal detectors using the generalized likelihood ratio test (GLRT) paradigm based on sample covariance matrices were proposed.  And in \cite{JieQijuan-MIMO},  the three high-performance detectors defined on egien-space of  sample covariance were proposed to achieve much better than conventional GLRT and  enegergy detection in terms of receiver operation curves (ROC). To improve the detection accuracy, the multi-layer neural networks (ML-NN) was introduced in \cite{LiYifanf-machine-learning} for inferring the number of passive emitters. Compared with the traditional signal detectors and classic information theoretic criteria like AIC and MDL, the ML-NN used the features extracted from the received signals for classification, and the final accuracy of inferring the number of signals reached more than 70\%.

In recent years, due to the fact the massive hybrid analog and digital (HAD) MIMO \cite{Large-scale-antenna-2014, SF-HAD-DOA-18, ZhangRuoyu-2021, LiSi-Covariance-Matrix-Reconstruction, Zhan-TLHAD-2022,Huang-hongji-18} owns three important properties:  low circuit loss, low computational complexity, and high-spatial-angle-resolution, DOA estimation based massive HAD MIMO  receiver has attracted a large amount of research activities. In \cite{SF-HAD-DOA-18}, three low-complexity and high-precision methods were developed to achieve the CRLB of HAD, but still require $M+1$ time-slots to eliminate the inherent deficiency of phase ambiguity, where $M$ is the number of antennas per sub-array. As $M$ increase, the DOA measurement delay will grow linearly. In order to reduce the estimate time delay, a fast phase ambiguity elimination method of finding the true emitter direction using only two time-slots was proposed in \cite{SBH-20}. In \cite{Zhan-TLHAD-2022}, a  two-layer HAD (TLHAD) structure was constructed, which can eliminate phase ambiguity with a single time-slot, which dramatically decreases the DOA measurement delay. Aiming at phase ambiguity of DOA estimation in large-scale HAD MIMO systems,
in \cite{Non-Circular}, an extended discrete fourier transform (DFT) algorithm is proposed, which combines the DFT method with non-circular (NC) signal and achieves DOA estimation using a single snapshot. Furthermore, in \cite{HuDie}, a low-complexity deep-learning  method  for hybrid HAD MIMO receiver using uniform circular arrays was proposed to learn the function of mapping the receiving signal vector into DOA.

Moreover, in \cite{LiSi-Covariance-Matrix-Reconstruction}, a beam sweeping approach was proposed to reconstruct spatial covariance matrix in hybrid massive MIMO systems such that MUSIC algorithm could be well applied to DOA estimation in massive HAD  MIMO. Subsequently, a machine-learning (ML) framework is proposed to improve the precision of measuring DOA in \cite{Zhuang-zhihong-20}. In \cite{Meng-Xiangming-18}, a generalized sparse Bayesian learning algorithm is integrated into the 1-bit DOA estimation. In \cite{SBH-ADC}, a new receive array framework with low-resolution ADCs was designed, and a closed-form expression of  CRLB was derived to evaluate the performance loss. From analysis, it is found that 2-3bit ADC is sufficient to achieve a trivial performance loss.  And, in \cite{shi-2021-mixed-ADCs}, the authors had made an investigation on the performance loss of DOA estimation under hybrid ADCs structure and show that low-resolution ADCs with only a few bits (up to 4 bits) can achieve an acceptable performance loss for DOA measurement.

As previously summarized, the two-layer hybrid structure has reduced $M$ time-slots in \cite{SF-HAD-DOA-18} and $2$ time-slots in \cite{SBH-20} to single one, which significantly improve the speed of DOA measurement. But this structure is made up of two distinct parts: fully-digital and hybrid sub-connected, which increase the complexity of designing RF circuit and leads to an increasing circuit cost. In this paper, only a pure  sub-connected hybrid structure is used to achieve a single-time-slot fast DOA estimator. This will make an effective complexity and cost reduction. Our main contributions are summarized as follows:
\begin{enumerate}
\item To rapidly eliminate the ambiguity of direction finding in HAD structure ,  a method of maximizing received power (Max-RP) is proposed. Its basic idea is to divide the angle range $[0,2\pi]$ into $K$ sectors. At receiver, analog beamforming per subarray is conducted  and is equal to the steering vector of its center angle of the corresponding sector. Finally, the sector center angle corresponding to the maximum value of all subarray outputs is just the true DOA. To further improve the estimate precision, a new methods is developed. The method is actually a quadratic interpolation using the three largest DOA values of Max-RP to improve its performance, called Max-RP-QI. In accordance with simulation results, the proposed two methods with approximately identical computational complexities have a descending order in performance: Max-RP-QI and Max-RP.
\item  To further enhance the estimate performance, a two-stage method is proposed as follows: generating a set of candidate solutions by Root-MUSIC and removing pseudo-solutions by  Max-RP-QI, called Root-MUSIC plus Max-RP-QI. Simulation results show that the proposed Root-MUSIC plus Max-RP-QI estimator can achieve the CRLB and performs much better than the proposed Max-RP-QI, and Max-RP in the high SNR regions with slightly higher-computational-complexity than Max-RP.
\end{enumerate}

The remainder of this paper is organized as follows. Section II describes the system model. In Section III, three estimators, Max-RP, Max-RP-QI and Root-MUSIC plus Max-RP-QI, are proposed to use only one time-slot to find the true direction angle of emitter at the cost of approximately $2dB\thicksim 4dB$ performance loss, and computational complexities are also analyzed. We present our simulation results in Section IV. Finally, we make our conclusions in Section V.

$Notations$: throughout this paper, boldface lower case and upper case letters represent vectors and matrices, respectively. Signs $(\cdot)^H$, $(\cdot)^{-1}$, and $\|\cdot\|$ denote the conjugate operation, conjugate transpose operation, inverse operation, and 2-norm operation, respectively. The notation $\textbf{I}_M$ is the $M\times M$ identity matrix. The sign $\mathbb{E}\{\cdot\}$ represents the expectation operation, $\textbf{diag}(\cdot)$ denotes the diagonal operator, $\arg(\cdot)$ means the argument of a complex number.

\section{system model}

\begin{figure}[h]
\centering
\includegraphics[width=0.35\textwidth]{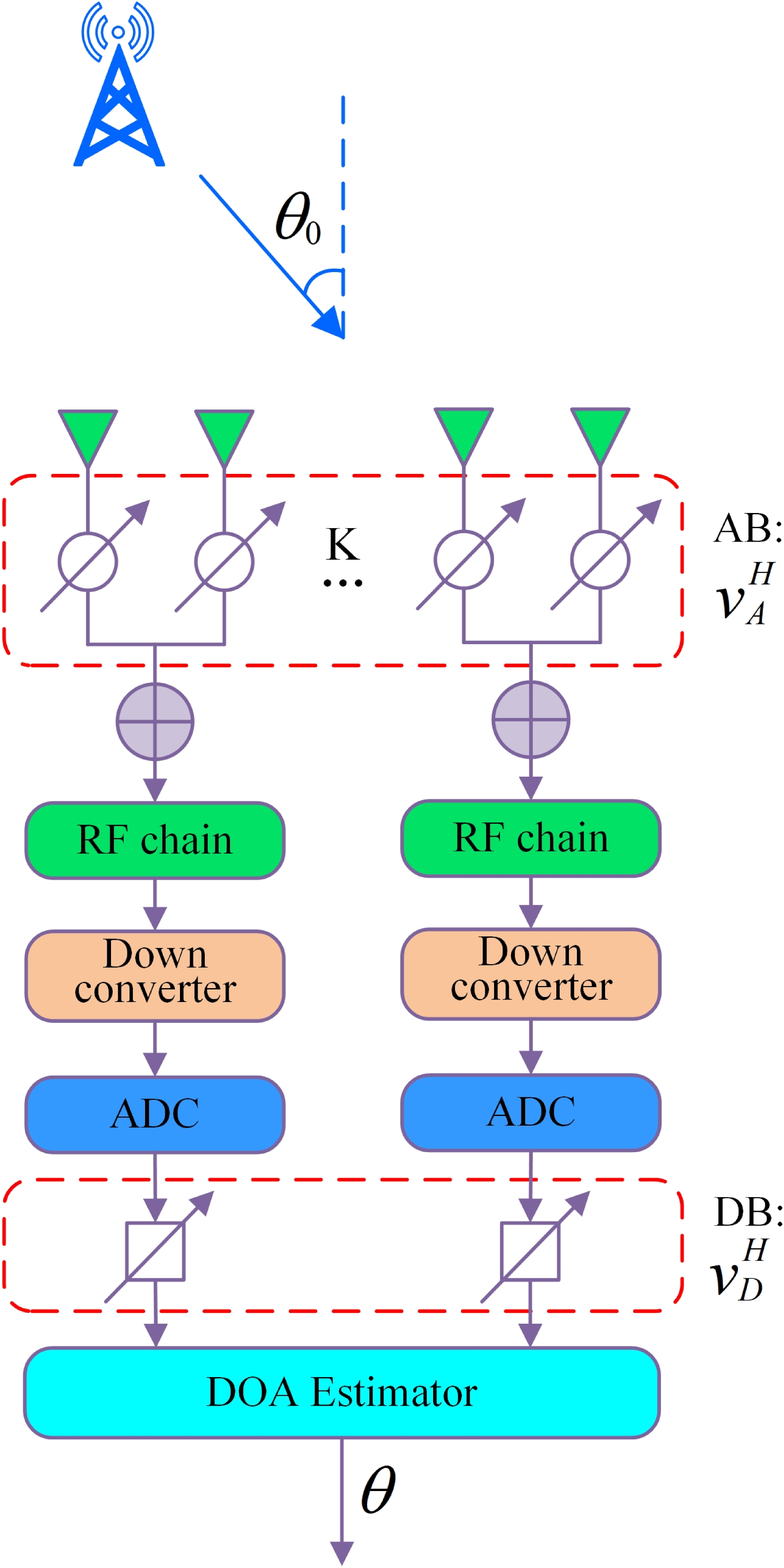}\\
\caption{ULA hybrid beamforming sub-connected architecture. (AB: analog beamforming, DB: digital beamforming)}\label{fig_system.eps}
\end{figure}

Fig. \ref{fig_system.eps} depicts the architecture of a HAD receiver based on massive MIMO. The HAD antenna array receives a far-field narrowband signal $s(t)e^{j2\pi f_c t}$, where $s(t)$ is the narrowband signal, $f_c$ is the carrier frequency, and  uniform linear array (ULA) with $N$ antennas. The ULA  is divided into $K$ subarrays with each subarry containing $M$ antennas, where $N=KM$.

Considering  sub-arrays are mutually independent of each other, the array  manifold  vector corresponding to subarray $k$ can be expressed as
\begin{align}
\textbf{a}_k(\theta_0)= e^{j2\pi \psi_k(\theta_0)}\left[e^{j2\pi \psi_{k}(1)},\cdots,e^{j2\pi\psi_{k}(M)}\right]^T
\end{align}
where $\psi_{k}(m)$ is the phase shift of signal at the baseband corresponding to the time delay from the source to antenna elements. And $\psi_{\theta_k}(m),m=1,2,...,M$ is expressed as
\begin{align}
\psi_{k}(\theta_0)=\frac{(k-1-K/2)Md\cos\theta_0}{\lambda}, 
\end{align}
and
\begin{align}
\psi_{k}(m)=\frac{(m-\frac{M}{2})d\cos\theta_0}{\lambda},m=1,2,...,M
\end{align}
where $\lambda$ is the wavelength of the carrier frequency. $d_m$ is the distance from a common reference point to the $m$th antenna. The  analog beamforming  (AB) vector corresponding to subarray $k$  is designed  as
\begin{align}
\textbf{v}_{A_k}(\theta_k)=e^{j2\pi \psi_k(\theta_0)}\textbf{a}^H_k(\theta_k)
\end{align}
such that phase alignment at subarray $k$ is achieved.

In this paper, $d$ is chosen to be half of the wavelength as usual, (i.e., $d=\frac{\lambda}{2}$). After passing through  parallel RF chain,  down-conversion and  ADC, the receive single $M$ dimensional vector of subarray $k$ from the emitter with direction angle being $\theta_0$ can be written as
\begin{align}\label{1}
\textbf{y}_k(t)=\textbf{v}_{A_k}^H(\theta_k)\textbf{a}(\theta_0)s(t)+\textbf{w}_k(t),k=1,2,..,K
\end{align}
where $\textbf{w}_k(t)\sim\mathcal{C}\mathcal{N}(0,\sigma^2_w\textbf{I}_M)$ is the additive white Gaussian noise (AWGN) vector. similarly, $\phi_{\theta_0}(m)=\frac{d_m\cos\theta_0}{\lambda},m=1,2,...,M$. And via digital beamforming (DB) operation, the receive single vector becomes
\begin{align}\label{1}
\textbf{r}_k(t)=\textbf{v}^H_{D_k}\textbf{v}_{A_k}^H(\theta_k)\textbf{a}(\theta_0)s(t)+\textbf{v}^H_{D_k}\textbf{w}_k(t)
\end{align}
where the DB vector is defined $\textbf{v}_{D_k} = [\upsilon_1, \upsilon_2,...,\upsilon_K]^T$. For convenience below, $\textbf{v}_{D_k} $ is chosen to be = $[1, 1,...,1]^T$.

\section{Proposed three fast DOA estimators}
In this section, to accelerate the elimination of phase ambiguity, the Max-RP method is first proposed, where the total angle range $[0,2\pi)$ are equally partitioned into $K$ sectors and each subarray takes charge of its owns sector. The center angle of the sector corresponding to the largest one among all subarray output powers is taken to be the final estimate DOA value. To improve the estimation accuracy, following Max-RP, a  quadratic interpolation  was performed by using the three largest RPs, called Max-RP-QI. To further enhance the estimate performance close to CRLB, the combining method of Root-MUSIC and Max-RP-QI was designed, and contains the following two steps: generating a set of candidate solutions by Root-MUSIC and removing pseudo-solutions by Max-RP-QI.
\subsection{Proposed Max-RP}
Fig.~1 plots the block diagram of the  Max-RP proposed by us. In this figure, each subarray is in charge of a sample angle range corresponding to one sector. The total angle range $[0,2\pi]$ are divided into $K$ subintervals. The output RP is viewed a function of discrete subarray index. Max-RP is eqvilent to find the maximum of this function.

\begin{figure}[h]
\centering
\includegraphics[width=0.35\textwidth]{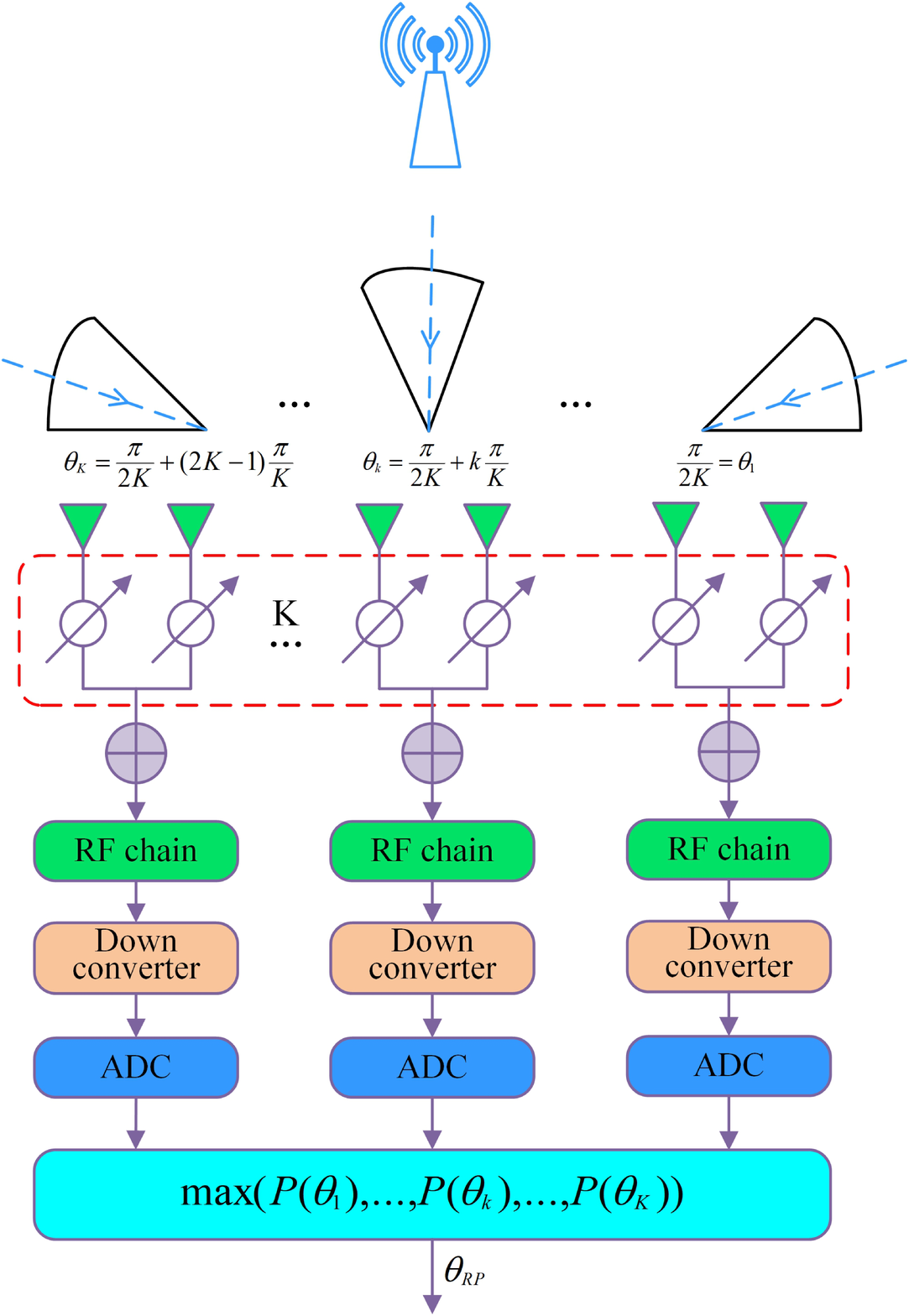}\\
\caption{Proposed Max-RP structure}\label{fig_maxp.eps}
\end{figure}
From  Fig.~1 and (\ref{1}),  after AB,  the  output of subarray  $k$ is
\begin{align}
y_k(t)=\gamma_k(\theta_k-\theta_0)s(t)+w_k(t),k=1,2,..,K
\end{align}
where
\begin{align}\label{2}
\gamma_k(\theta_k-\theta_0)&=\textbf{v}_{A_k}^H(\theta_k)\textbf{a}(\theta_0)\nonumber\\
&=\frac{\exp\Big\{j\frac{2\pi }{\lambda}Md(\cos\theta_0-\cos\theta_k)\Big\}-1 }{ \exp\Big\{j\frac{2\pi }{\lambda}d(\cos\theta_0-\cos\theta_k)\Big\}-1}
\end{align}
with $\theta_k$ belonging to
\begin{align}\label{2}
\Theta=\left\{\frac{\pi}{K},\frac{3\pi}{K},\cdot, \frac{(2K-1)\pi}{K}\right\}
\end{align}
It is assumed that the beamforming vector is known, the received signal power can be expressed as
\begin{align}\label{3}
&P(\theta_k)=\mathbb{E}\left\{y_k(t)y_k(t)^H\right\}\nonumber\\&=\mathbb{E}\Big\{\big[\gamma_k(\theta_k-\theta_0)s(t)+w_k(t)\big]\big[\gamma_k(\theta_k-\theta_0)s(t)+w_k(t)\big]^H\Big\}
\end{align}
Since noise is independent and uncorrelated, we have
\begin{subequations}\label{4}
\begin{align}
&\mathbb{E}\big\{\gamma_k(\theta_k-\theta_0)s(t)w_k(t)^H\big\}\nonumber\\&=\gamma_k(\theta_k-\theta_0)\mathbb{E}\big\{s(t)w_k(t)^H\big\}=0\\
&\mathbb{E}\big\{w_k(t)\big[\gamma_k(\theta_k-\theta_0)s(t)\big]^H\big\}\nonumber\\&=\gamma_k(\theta_k-\theta_0)^*\mathbb{E}\big\{w_k(t)s(t)^H\big\}=0
\end{align}
\end{subequations}
Substituting (\ref{4}) into (\ref{3}) yields
\begin{align}
P(\theta_k)&=\mathbb{E}\left\{y_k(t)y_k(t)^H\right\}\nonumber\\
&=\gamma_k(\theta_k-\theta_0)\mathbb{E}\left\{s(t)s(t)^H\right\}\gamma_k(\theta_k-\theta_0)^*\nonumber\\&+\mathbb{E}\left\{w_k(t)w_k(t)^H\right\}
\end{align}
where $P_{s,k}=\mathbb{E}\{s(t)s(t)^H\}$ and $\sigma^2_{w,k}=\mathbb{E}\{w_k(t)w_k(t)^H\}$. Then, we have
\begin{align}
P(\theta_k)=P_{s,k}+\sigma^2_{w,k},k=1,2,...,K
\end{align}
Actually, $P(\theta_k)$ cannot be obtained directly. However, it can be obtained from the available data and can be expressed as
\begin{align}\label{5}
&P(\theta_k)\nonumber\\
&=\frac{1}{L}\sum^L_{l=1}\left\{y_k(l)y_k(l)^H\right\}\nonumber\\
&=\frac{1}{L}\sum^L_{l=1}\Big\{\big[\gamma_k(\theta_k-\theta_0)s(l)+w_k(l)\big]\nonumber\\&~~~~~~~~~~~~~~\big[\gamma_k(\theta_k-\theta_0)s(l)+w_k(l)\big]^H\Big\}\nonumber\\
&=\frac{1}{L}\Big(\sum^L_{l=1}\big\{\big[\gamma_k(\theta_k-\theta_0)s(l)\big]\big[\gamma_k(\theta_k-\theta_0)s(l)\big]^H\big\}\nonumber\\
&+\sum^L_{l=1}\big\{\big[\gamma_k(\theta_k-\theta_0)s(l)\big]w_k(l)^H+w_k(l)\big[\gamma_k(\theta_k-\theta_0)s(l)\big]^H\big\}\nonumber\\&+\sum^L_{l=1}\big\{w_k(l)w_k(l)^H\big\}\Big)
\end{align}
Similarly, the noise is independent and uncorrelated, we have
\begin{align}
&\sum^L_{l=1}\big\{\big[\gamma_k(\theta_k-\theta_0)s(l)\big]w_k(l)^H+w_k(l)\big[\gamma_k(\theta_k-\theta_0)s(l)\big]^H\big\}\nonumber\\&=0
\end{align}
It follows from the above that (\ref{5}) is rewritten as
\begin{align}\label{6}
P(\theta_k)=&\frac{1}{L}\Big(\sum^L_{l=1}\big[\gamma_k(\theta_k-\theta_0)s(l)\big]\big[\gamma_k(\theta_k-\theta_0)s(l)\big]^H\big\}\nonumber\\&+\sum^L_{l=1}\big\{w_k(l)w_k(l)^H\big\}\Big)
\end{align}
Let us defined $P_s=\frac{1}{L}\sum^L_{l=1}\{s(l)s(l)^H\}$ and $\sigma^2_w=\frac{1}{L}\sum^L_{l=1}\{w_k(l)w_k(l)^H\}$. Due to the $\gamma(\theta_k-\theta_0)$ is a scalar, so when $L$ is sufficiently large, we have
\begin{align}
&\frac{1}{L}\sum^L_{l=1}\left\{\big[\gamma_k(\theta_k-\theta_0)s(l)\big]\big[\gamma_k(\theta_k-\theta_0)s(l)\big]^H\right\}\nonumber\\
&=P_s\gamma(\theta_k-\theta_0)\gamma(\theta_k-\theta_0)^*=P_s\gamma(\theta_k-\theta_0)^2
\end{align}
It follows from the above that (\ref{6}) is rewritten as
\begin{align}
P(\theta_k)=P_s\gamma(\theta_k-\theta_0)^2+\sigma^2_w
\end{align}

The received signal power of this receiver structure can therefore be expressed in the vector form as follows
\begin{align}
\textbf{P}=\left[P(\theta_1),P(\theta_2),...,P(\theta_K)\right]
\end{align}

Eventually, the estimated direction angle  is
\begin{align}
\theta_{RP}=\arg \max \limits_{\boldsymbol{\theta_k\in\Theta}} \textbf{P}(\theta_k)
\end{align}
\begin{figure}[h]
\centering
\includegraphics[width=0.40\textwidth]{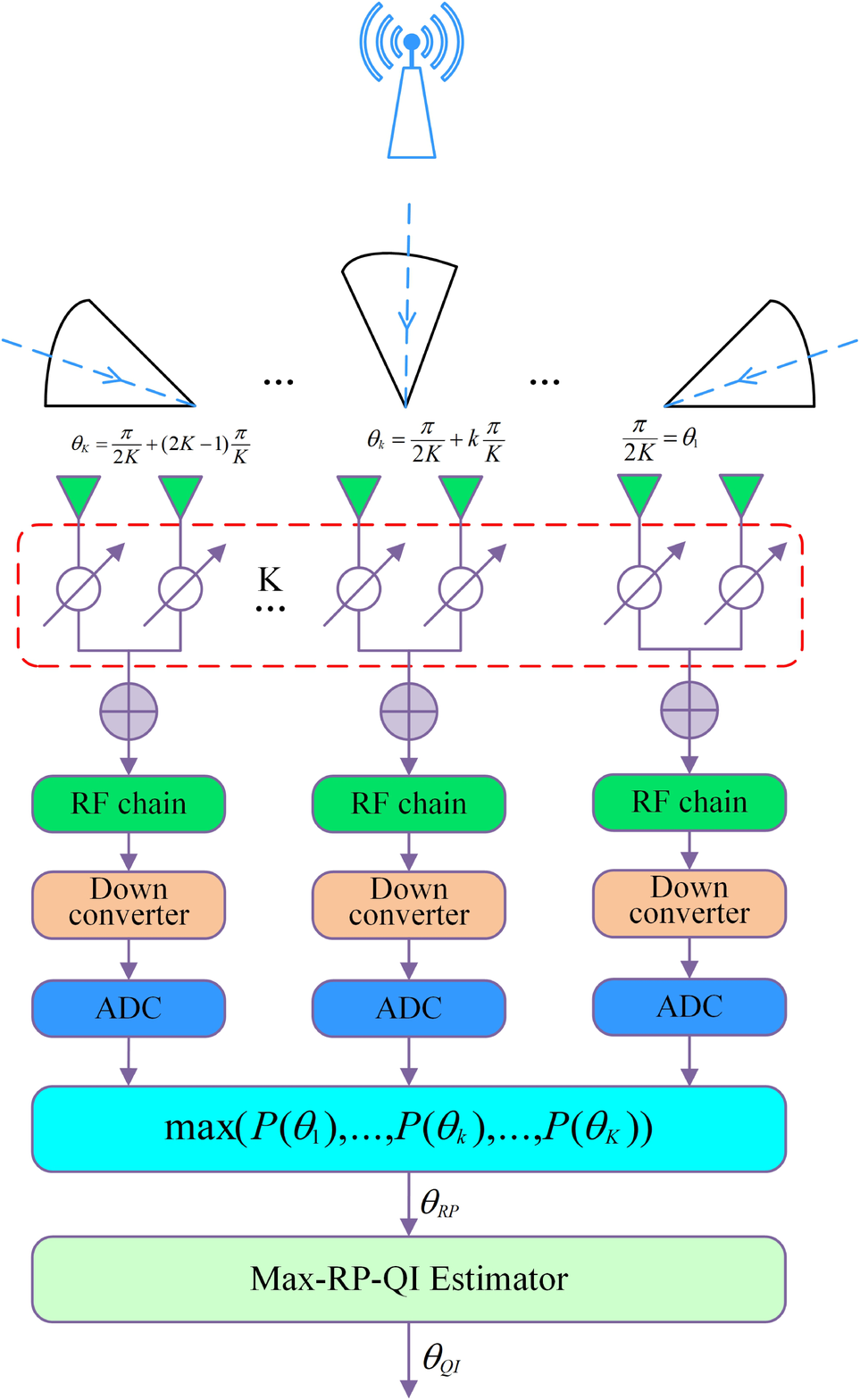}\\
\caption{Proposed Max-RP-QI structure}\label{fig_qi.eps}
\end{figure}

\subsection{Proposed Max-RP-QI}
In the preceding subsection, the Max-RP estimator is actually a global search with stepsize $2\pi/K$, and its resolution is $2\pi/K$. This will lead to a result that improving its precision require to increase the number of subarrays. To improve its estimate accuracy, following the maximizing operation, a quadratic interpolation is used over the three largest RPs.  The new quadratic curve is maximized to find the more precise DOA.

From \cite{Matrix-01},  a quadratic form is given as
\begin{align}\label{7}
f(\theta)=c+b\theta+a\theta^2 =P(\theta)
\end{align}
Taking the derivative of the above (\ref{7}) with respect to $\theta$ equals zero, we have
\begin{align}
\theta_{QI}=-\frac{b}{2a}
\end{align}

According to Fig.~\ref{fig_qi.eps}, assuming that the received signal power achieves its maximum value in the $k$th sector, the system of linear equations of variables $a$, $b$ and $c$ is expressed as
\begin{equation}\label{8}
\left\{
\begin{array}{lr}
P(\theta_{k-1})=c+b\theta_{k-1}+a\theta^2_{k-1}&\\
P(\theta_{k})=c+b\theta_{k}+a\theta^2_{k},  &k=1,2,...,K\\
P(\theta_{k+1})=c+b\theta_{k+1}+a\theta^2_{k+1}&\\
\end{array}
\right.
\end{equation}
where $\theta_{k}=\hat{\theta}$ corresponds  to the subarray or sector index of the maximum RP.  The equations (\ref{8}) is rewritten in the matrix-vector form
\begin{equation}
\left[
\begin{array}{cccc}
 1& \theta_{k-1} & \theta^2_{k-1}\\
 1& \theta_{k} & \theta^2_{k}\\
 1& \theta_{k+1} &\theta^2_{k+1}
\end{array}
\right ]
\left[
\begin{array}{cccc}
 c\\
 b\\
 a
\end{array}
\right ]
=
\left[
\begin{array}{cccc}
P(\theta_{k-1})\\
P(\theta_{k})\\
P(\theta_{k+1})
\end{array}
\right ]
\end{equation}
Then, let us define
\begin{align}
\textbf{A}=\begin{bmatrix}
 1& \theta_{k-1} & \theta^2_{k-1}\\
 1& \theta_{k} & \theta^2_{k}\\
 1& \theta_{k+1} &\theta^2_{k+1}
\end{bmatrix}
\end{align}
and
\begin{align}
\textbf{x}=\begin{bmatrix}
 c\\
 b\\
 a
\end{bmatrix}~~~~~
\textbf{b}=\begin{bmatrix}
P(\theta_{k-1})\\
P(\theta_{k})\\
P(\theta_{k+1})
\end{bmatrix}
\end{align}
Then
\begin{align}
\textbf{b}=\textbf{A}\textbf{x}
\end{align}

Considering $\textbf{A}$ is a Vandermonde  matrix and $\theta_{k+1}\neq\theta_{k}\neq\theta_{k-1}$ , $\det{\textbf{A}}\neq0$ and $\textbf{A}$ is invertible, it has a unique inverse matrix, denoted as $\textbf{A}^{-1}$. Then,
\begin{align}
\textbf{x}=\textbf{A}^{-1}\textbf{b}
\end{align}
Since  $\textbf{A}$ is a Vandermonde  matrix,  using Lagrangian interpolation, we directly have (\ref{A}), (\ref{b}), and  (\ref{a}).
\begin{figure*}[h]
\begin{equation}\label{A}
\textbf{A}^{-1}=
\left[
\begin{array}{ccc}
 \frac{(-1)^{2}\theta_{k}\theta_{k+1}}{(\theta_{k-1}-\theta_{k})(\theta_{k-1}-\theta_{k+1})}& \frac{(-1)^{2}\theta_{k-1}\theta_{k+1}}{(\theta_{k}-\theta_{k-1})(\theta_{k}-\theta_{k+1})} & \frac{(-1)^{2}\theta_{k-1}\theta_{k}}{(\theta_{k+1}-\theta_{k-1})(\theta_{k+1}-\theta_{k})}\\
 \frac{(-1)(\theta_{k}+\theta_{k+1})}{(\theta_{k-1}-\theta_{k})(\theta_{k-1}-\theta_{k+1})}& \frac{(-1)(\theta_{k-1}+\theta_{k+1})}{(\theta_{k}-\theta_{k-1})(\theta_{k}-\theta_{k+1})} & \frac{(-1)(\theta_{k-1}+\theta_{k})}{(\theta_{k+1}-\theta_{k-1})(\theta_{k+1}-\theta_{k})}\\
 \frac{1}{(\theta_{k-1}-\theta_{k})(\theta_{k-1}-\theta_{k+1})}& \frac{1}{(\theta_{k}-\theta_{k-1})(\theta_{k}-\theta_{k+1})} &\frac{1}{(\theta_{k+1}-\theta_{k-1})(\theta_{k+1}-\theta_{k})}
\end{array}
\right ]
\end{equation}
\end{figure*}
\begin{figure*}[h]
\begin{align}\label{b}
b=&\frac{(-1)(\theta_{k}+\theta_{k+1})P(\theta_{k-1})}{(\theta_{k-1}-\theta_{k})(\theta_{k-1}-\theta_{k+1})}
+\frac{(-1)(\theta_{k-1}+\theta_{k+1})P(\theta_{k})}{(\theta_{k}-\theta_{k-1})(\theta_{k}-\theta_{k+1})}
+\frac{(-1)(\theta_{k-1}+\theta_{k})P(\theta_{k+1})}{(\theta_{k+1}-\theta_{k-1})(\theta_{k+1}-\theta_{k})}\nonumber\\=&
\frac{(\theta^2_{k-1}-\theta^2_{k+1})P(\theta_{k})-(\theta^2_{k-1}-\theta^2_{k})P(\theta_{k+1})-(\theta^2_{k}-\theta^2_{k+1})P(\theta_{k-1})}{(\theta_{k-1}-\theta_{k})(\theta_{k-1}-\theta_{k+1})(\theta_{k}-\theta_{k+1})}\nonumber\\=&
\frac{(\theta^2_{k}-\theta^2_{k+1})(P(\theta_{k})-P(\theta_{k-1}))-(\theta^2_{k-1}-\theta^2_{k})(P(\theta_{k})-P(\theta_{k+1}))}{(\theta_{k-1}-\theta_{k})(\theta_{k-1}-\theta_{k+1})(\theta_{k}-\theta_{k+1})}
\end{align}
\end{figure*}
\begin{figure*}[h]
\begin{align}\label{a}
a=&\frac{P(\theta_{k-1})}{(\theta_{k-1}-\theta_{k})(\theta_{k-1}-\theta_{k+1})}
+\frac{P(\theta_{k})}{(\theta_{k}-\theta_{k-1})(\theta_{k}-\theta_{k+1})}
+\frac{P(\theta_{k+1})}{(\theta_{k+1}-\theta_{k-1})(\theta_{k+1}-\theta_{k})}\nonumber\\=&
\frac{(\theta_{k}-\theta_{k+1})P(\theta_{k-1})-(\theta_{k-1}-\theta_{k+1})P(\theta_{k})+(\theta_{k-1}-\theta_{k})P(\theta_{k+1})}{(\theta_{k-1}-\theta_{k})(\theta_{k-1}-\theta_{k+1})(\theta_{k}-\theta_{k+1})}\nonumber\\=&
\frac{(\theta_{k}-\theta_{k-1})(P(\theta_{k})-P(\theta_{k+1}))-(\theta_{k}-\theta_{k+1})(P(\theta_{k})-P(\theta_{k-1}))}{(\theta_{k-1}-\theta_{k})(\theta_{k-1}-\theta_{k+1})(\theta_{k}-\theta_{k+1})}
\end{align}
\end{figure*}
Then,we have (\ref{theta}).
\begin{figure*}[h]
\begin{align}\label{theta}
&\theta_{QI}=-\frac{1}{2}\times\frac{(\theta^2_{k}-\theta^2_{k+1})(P(\theta_{k})-P(\theta_{k-1}))-(\theta^2_{k-1}-\theta^2_{k})(P(\theta_{k})-P(\theta_{k+1}))}{(\theta_{k}-\theta_{k-1})(P(\theta_{k})-P(\theta_{k+1}))-(\theta_{k}-\theta_{k+1})(P(\theta_{k})-P(\theta_{k-1}))}
\end{align}
\rule[-10pt]{18.3cm}{0.05em}
\end{figure*}
\subsection{Proposed Root-MUSIC plus Max-RP-QI}
\begin{figure}[h]
\centering
\includegraphics[width=0.40\textwidth]{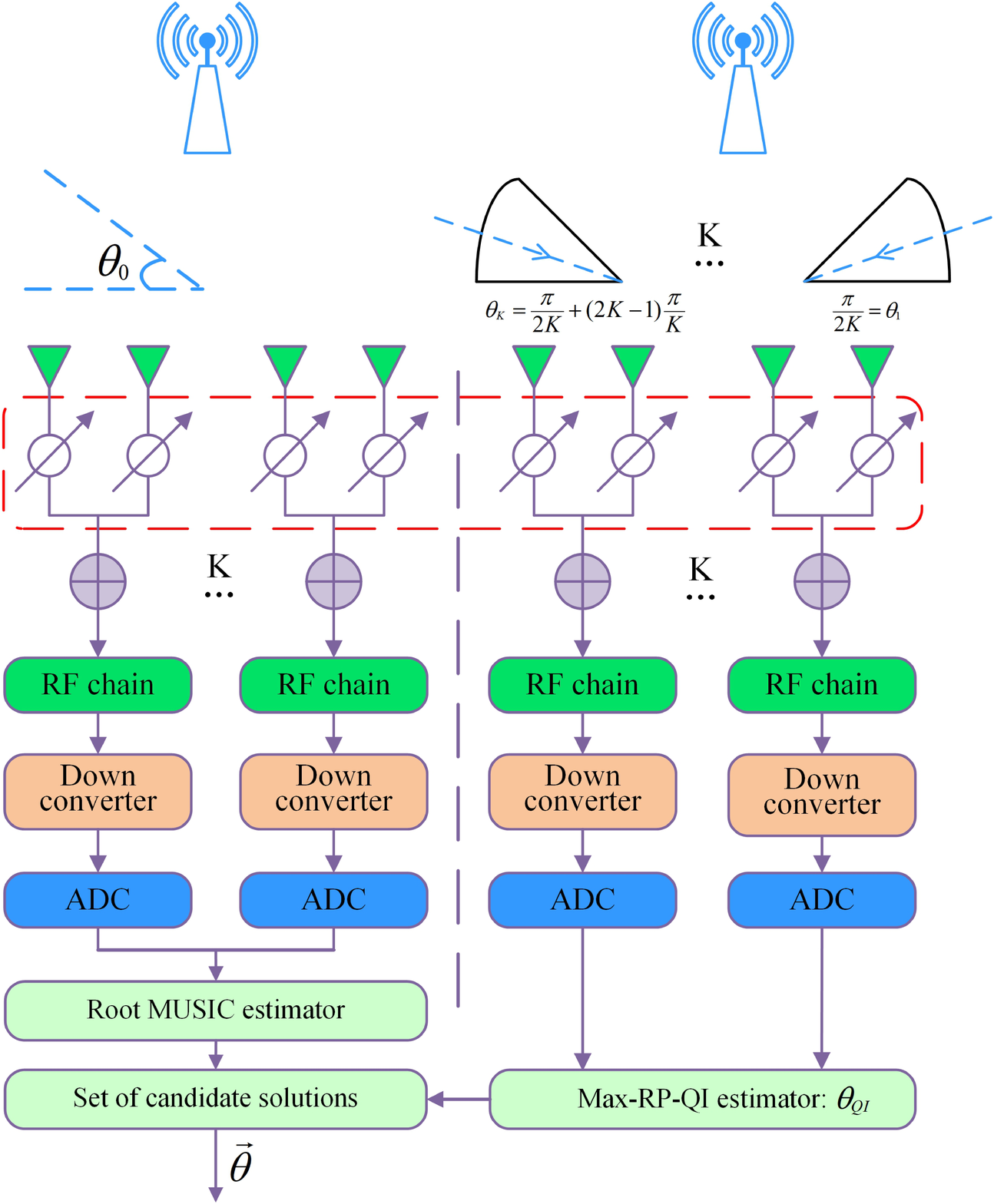}\\
\caption{Proposed Root-MUSIC Plus Max-RP-QI structure}\label{fig_root.eps}
\end{figure}

In the section, to achieve the CRLB, the Root-MUSIC \cite{Root-Music-1989, Root-MUSIC-1993} is combined with the proposed Max-RP-QI to form a blended scheme: Root-MUSIC plus Max-RP-QI as shown in Fig.~\ref{fig_root.eps}.  The left part of this array structure uses  a small-scale number $Q$ of subarrays, totally $N_a=MQ$ antennas,  to generate a set of candidate solutions. For the right part of this HAD structure, the Max-RP-QI is adopted to output the optimal DOA, which is to choose the true DOA value from the set of candidate solutions.

 Given the initial phases of all analog phase shifters of the left part in Fig.~\ref{fig_root.eps} are zeros, we have
\begin{align}\label{9}
\textbf{V}_A=\frac{1}{\sqrt{M}}[1,1,...,1]^T
\end{align}
According to (\ref{9}), the receive vector consisting of all outputs of the subarray in the left part is expressed as
\begin{align}\label{R-Signal-Vec-L}
\textbf{y}_L(t)&=[y_1(t),y_2(t),...,y_q(t)]^T\nonumber\\&=
\frac{1}{\sqrt{M}}\textbf{a}_Q(\theta_0)g(\theta_0)s(t)+[w_1(t),w_2(t),...,w_Q(t)]^T
\end{align}
where $\textbf{a}_Q(\theta_0)$ is the associated small-scale array manifold  written as
\begin{align}\label{11}
\textbf{a}_Q(\theta_0)=\left[e^{j2\pi\mu_{\theta_0}(1)},e^{j2\pi\mu_{\theta_0}(2)},..,e^{j2\pi\mu_{\theta_0}(Q-1)}\right]^T
\end{align}
where
\begin{align}
\mu_{\theta_0}(q)=\frac{Md_q\sin\theta_0}{\lambda},q=1,2,...,Q-1.
\end{align}

And $g(\theta_0)$ in (\ref{R-Signal-Vec-L}) is the constant obtained by summing the elements of each subarray oriented vector can be denoted as
\begin{align}\label{12}
g(\theta_0)&=\sum_{m=1}^{M}e^{j\frac{2\pi}{\lambda}(m-1)d\sin\theta_0}=\frac{1-e^{j\frac{2\pi}{\lambda}Md\sin\theta_0}}{1-e^{j\frac{2\pi}{\lambda}d\sin\theta_0}}
\end{align}
Therefore, the array manifold $\textbf{a}_y(\theta_0)$ is defined as
\begin{align}
\textbf{a}_y(\theta_0)=\textbf{a}_Q(\theta_0)g(\theta_0)
\end{align}

From (\ref{R-Signal-Vec-L}), the covariance matrix of the output vector is defined as
\begin{align}\label{13}
\textbf{R}_y&=\mathbb{E}[\textbf{y}\textbf{y}^H]^T\nonumber\\&=\textbf{a}_y\textbf{R}_s\textbf{a}_y^H+\textbf{R}_w
\end{align}
Subsitting (\ref{11}) and (\ref{12}) into (\ref{13}), the above formula is rewritten as
\begin{align}\label{13}
\textbf{R}_y=\frac{1}{M}\sigma_s^2\|g(\theta_0)\|^2\textbf{a}_Q(\theta_0)\textbf{a}^H_Q(\theta_0)+\sigma^2_w\textbf{I}
\end{align}
where $\sigma_s^2$ represents the variance of the receive signal being the average receive signal power and $\sigma_s^2$ stands for the noise variance.  Furthermore, similar to the conventional Root-MUSIC method, the eigenvalue decomposition (EVD) of $\textbf{R}_y$ is expressed as
\begin{align}
\textbf{R}_{y}&=\textbf{U}\bf{\Sigma}\textbf{U}^H\nonumber\\
&=[\textbf{U}_S\,\textbf{U}_N]\bf{\Sigma}[\textbf{U}_S\,\textbf{U}_N]^H
\end{align}
where
\begin{equation}
\bf{\Sigma}=
\left[
\begin{array}{cccc}
 \sigma_s^2+\sigma^2_w& 0 &\cdots  & 0\\
 0& \sigma^2_w &\cdots& 0\\
 \vdots  & \vdots & \ddots & \vdots \\
 0& 0 &\cdots  &\sigma^2_w
\end{array}
\right ]
\end{equation}
is a diagonal matrix with elements being the corresponding eigen-values of $\textbf{R}_{y}$. And the $Q\times 1$ matrix $\textbf{U}_S$ contains the singular vectors corresponding to the largest eigen-value, the matrix $\textbf{U}_N$ contains the eigen-vectors corresponding to the $Q-1$ smallest singular values. Then, we compute
\begin{align}\label{14}
\varphi(\theta)&=\frac{1}{\|\textbf{U}^H_N\textbf{a}_y^H\|^2}=\frac{1}{\|g(\theta_0)\|^2\|\textbf{a}^H_Q\textbf{U}_N\textbf{U}^H_N\textbf{a}_Q\|}
\end{align}
which will have peaks at the signal directions.

In order to obtain the peak point of (\ref{14}), the Root-MUSIC algorithm is adopted because it does not require linear search, has low complexity and can obtain approximate analytical solutions. Then, the polynomial $f(z)$ is defined as
\begin{align}
f(z)=g^H(z)\textbf{a}^H_Q(z)\textbf{U}_N\textbf{U}^H_N\textbf{a}_Q(z)g(z)=0
\end{align}
where
\begin{align}
z=e^{j\frac{2\pi}{\lambda}Md\sin\theta}
\end{align}

Obviously, the order of the polynomial is $2(Q-1)$, which means that the polynomial has $2(Q-1)$ roots. Then, for an equally spaced uniform linear array, the set of the candidate solutions can be given by
\begin{align}
{\Theta}_z=\Big\{\tilde{\theta}_1,\tilde{\theta}_2,\cdots, \tilde{\theta}_{2Q-2}\Big\}
\end{align}
where
\begin{align}
\tilde{\theta}_q=\arcsin\left(\frac{\lambda\arg z_q}{2\pi Md}\right)
\end{align}
After estimating the DOA value $\theta_{QI}$ using the right-hand structure, the optimal solution can be written as
\begin{align}
\overrightarrow{\theta}_=\arg \min \limits_{\theta_q \in \Theta_z}\|\tilde{\theta}_q-\overline{\theta}_{QI}\|^2
\end{align}
This completes the design of the proposed Root-MUISC plus Max-RP-QI.

\subsection{Complexity Analysis}
In what follows, we make an analysis of computational complexities of the proposed three estimators with previous TLHAD method as a complexity reference. Thus, the
complexity of Max-RP is as follows
\begin{align}
C_{Max-RP}=O\Big\{K(3ML^2+L^2-M)\Big\}
\end{align}
FLOPs. The complexity of Max-RP-QI is
\begin{align}
C_{Max-RP-QI}=O\Big\{&K(3ML^2+L^2-M)\nonumber\\&+\frac{1}{3}K^3+2K^2+\frac{2}{3}K\Big\}
\end{align}
FLOPs. The complexity of Root-MUSIC Plus Max-RP-QI is
\begin{align}
&C_{Root-MUSIC Plus Max-RP-QI}=O\Big\{(K-2)\times\nonumber\\&\big((3ML^2+L^2-M)+\frac{1}{3}K^3+2K^2+\frac{2}{3}K\big)+M^3+2ML^2\Big\}
\end{align}
FLOPs. And the complexity of the existing TLHAD is
\begin{align}
C_{TLHAD}=O\Big\{M^3+M^2(2L-1)+\frac{N^3}{4}+\frac{N^2}{4}(2L-1)\Big\}
\end{align}
FLOPs. Considering the number of antennas tends to large-scale or ultra-large-scale, compare with the previous TLHAD estimators, the computational complexities of the proposed three estimators are significantly reduced.
\section{Numerical results}
In this section, we present simulation results to the performance of the three estimators: Max-RP, Max-RP-QI and Root-MUSIC Plus Max-RP-QI with the hybrid CRLB as a performance benchmak. Simulation parameters are chosen as shown: the direction of emitter $\theta_0 = 41^o$, and $L \in \{50,100,200,400\}$. In large-scale and ultra-large-scale MIMO scenarios, the number of antennas at receive array is set $N=1024$ and $M=8$.

\begin{figure}[h]
\centering
\includegraphics[width=0.5\textwidth]{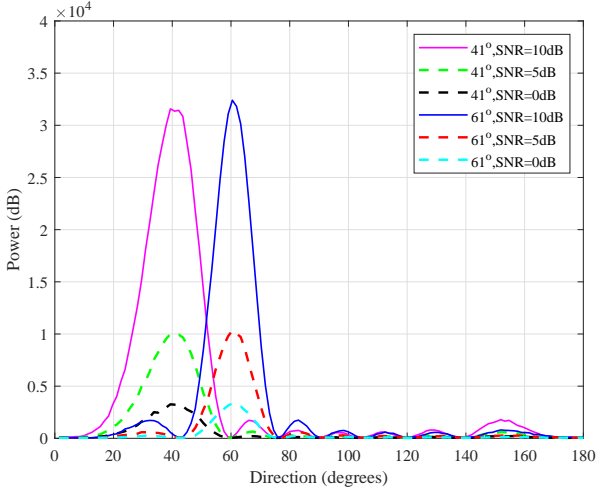}\\
\caption{Receive power versus direction estimated by the proposed Max-RP method}\label{fig_p.eps}
\end{figure}

Fig.~\ref{fig_p.eps} demonstrates the receive power  versus DOA estimated by Max-RP  for $N=1024$, $K=128$ , and $\theta_0 \in \{41^o, 61^o\}$. As can be seen, $P(\theta_k)$ is maximized around  the source direction $\theta$ for three different SNRs (0dB, 5dB, 10dB). And we find that the proposed Max-RP estimators can provide a prefect estimate of direction for $\text{SNR} \geq 5\text{dB}$.

\begin{figure}[h]
\centering
\includegraphics[width=0.5\textwidth]{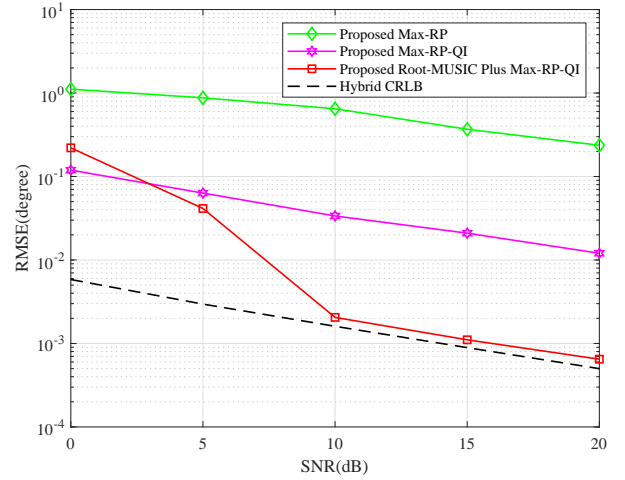}\\
\caption{RMSE versus SNR of the proposed methods}\label{fig_rmse.eps}
\end{figure}

Fig.~\ref{fig_rmse.eps} plots the  root mean squared error (RMSE) versus SNR of the proposed three methods, where the corresponding hybrid CRLBs are used as a performance benchmark. From this figure, it is seen that the proposed Root-MUSIC plus Max-RP-QI method can achieve the CRLBs with the $\text{SNR} \geq 10\text{dB}$ while  Max-RP and Max-RP-QI are still  substantial gaps from the hybrid CRLBs.

\begin{figure}[h]
\centering
\includegraphics[width=0.5\textwidth]{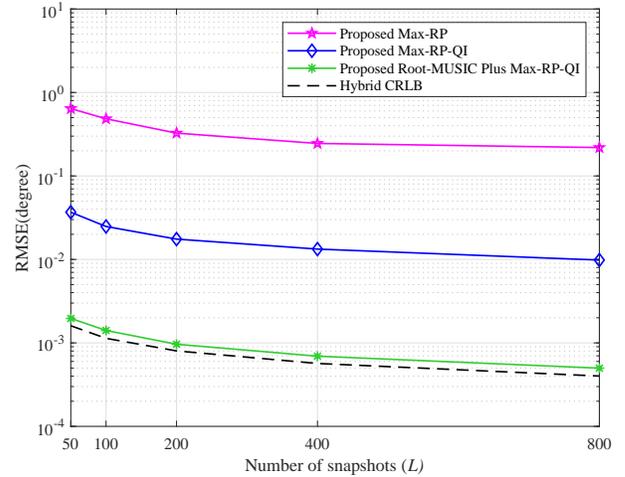}\\
\caption{RMSE versus snapshots of the proposed methods}\label{fig_rmse_L.eps}
\end{figure}

To observe the impact of the different $L$ of number of the snapshots on the proposed schemes, in Fig.~\ref{fig_rmse_L.eps},  the value of $L$ is varied from 50 to 400, given the number of antennas N = 1024, and the number of sunarray $K$ = 128. Thus, Fig.~\ref{fig_rmse_L.eps}  plots the RMSE verus the number of snapshots of the proposed method. From this figure, we obtain the same performance trend as Fig.~\ref{fig_rmse.eps}. Additionally,  it is noted that, as the number $L$ of snapshots increases, the accuracy of the proposed Root-MUSIC plus Max-RP-QI methods is accordingly improved.

\begin{figure}[h]
\centering
\includegraphics[width=0.5\textwidth]{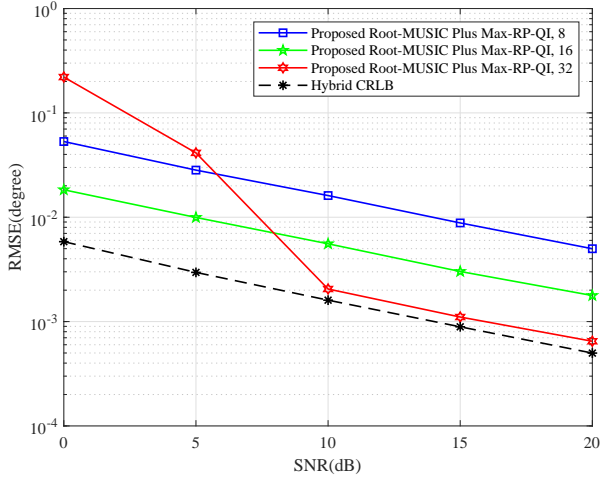}\\
\caption{RMSE versus SNR of the proposed Root-MUSIC Plus Max-RP-QI method for different number of subarrays $K$}\label{fig_root_rmse.eps}
\end{figure}

To evaluate the impact of the number of assigned subarrays of the left part structure on perfromance, Fig.~\ref{fig_root_rmse.eps} plots  the curves of RMSE versus SNR for  $K_L$=8, 16,  and 32, where $K_L$ denotes three different numbers of left-side subarrays for Root-MUSIC. Observing this figure, we find that the proposed Root-MUSIC Plus Max-RP-QI estimator can achieve the hybrid CRLB for $K_L=32$.

\begin{figure}[h]
\centering
\includegraphics[width=0.5\textwidth]{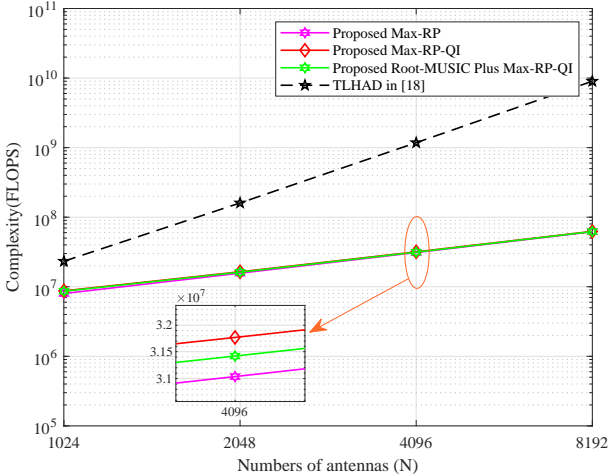}\\
\caption{Complexity versus number of antennas of the proposed methods}\label{complexity.eps}
\end{figure}

Fig.~\ref{complexity.eps} illustrates the curves of complexity versus the number $N$ of antennas with $N$ varying from 1024 to 8192. From Fig.~\ref{complexity.eps}, it is shown that the proposed three methods have  approximately the same computational complexities and far lower than existing TLHAD  in \cite{Zhan-TLHAD-2022}.


\section{Conclusions}
In this paper, based on the hybrid massive MIMO receive array structure, the three DOA estimators: Max-RP, Max-RP-QI, and Root-MUSIC Plus Max-RP-QI, were poposed, which succefully  eliminate phase ambiguity in a single time-slot. In summary, the proposed methods have an increasing precision order as follows: Max-RP, Max-RP-QI, and Root-MUSIC Plus Max-RP-QI.  Moreover, the proposed Root-MUSIC Plus Max-RP-QI can achieve CRLB with  approximately identical computational complexity as the first two methods. This makes them attractive for the future applications of DOA-measurement based sensing in IoT, UAV, satellite communications, WSNs, DM network, and beyond 5G.
\ifCLASSOPTIONcaptionsoff
  \newpage
\fi

\bibliographystyle{IEEEtran}
\bibliography{IEEEfull,reference}
\end{document}